\documentclass[ aip,amsmath,amssymb,reprint]{revtex4-1}

\usepackage[utf8]{inputenc}
\usepackage[T1]{fontenc}
\usepackage{mathptmx}
\usepackage{comment}
\usepackage{color}

\usepackage{graphicx}
\usepackage{dcolumn}
\usepackage{txfonts}
\usepackage{microtype}
\usepackage{bm}
\usepackage{ulem}
\usepackage{mathrsfs}
\usepackage[export]{adjustbox}
\usepackage{epstopdf}

\begin{document}

%\preprint{AIP/123-QED}

\title{Photon polarization effects in polarized electron-positron pair production in a strong laser field}

\author{Ya-Nan Dai}
\affiliation{Department of Physics, Shanghai Normal University, Shanghai 200234, China}
%\author{Yue-Yue Chen}\email{yueyuechen@shnu.edu.cn}\affiliation{Department of Physics, Shanghai Normal University, Shanghai 200234, China}
\author{Bai-Fei Shen}
\affiliation{Department of Physics, Shanghai Normal University, Shanghai 200234, China}
\author{Jian-Xing Li}
\affiliation{School of Physics, Xi’an Jiaotong University, Xi’an 710049, China}
\author{Rashid Shaisultanov}
\affiliation{Max-Planck-Institut fur Kernphysik, Saupfercheckweg 1, 69117 Heidelberg, Germany}
\affiliation{Helmholtz-Zentrum Dresden-Rossendorf, Bautzner Landstraße 400, 01328 Dresden, Germany}
\author{Karen Z. Hatsagortsyan}
%\email{k.hatsagortsyan@mpi-hd.mpg.de}
\affiliation{Max-Planck-Institut fur Kernphysik, Saupfercheckweg 1, 69117 Heidelberg, Germany}
\author{Christoph H. Keitel}
\affiliation{Max-Planck-Institut fur Kernphysik, Saupfercheckweg 1, 69117 Heidelberg, Germany}
\author{Yue-Yue Chen}
\email{yueyuechen@shnu.edu.cn}
\affiliation{Department of Physics, Shanghai Normal University, Shanghai 200234, China}

\date{\today}

\begin{abstract}

Deep understanding of photon polarization
%effects
impact on pair production is essential for the
%production
efficient creation of laser driven polarized positron beams,
%source, as well as
and demands a complete description of polarization effects in strong-field QED processes. We investigate, employing fully polarization resolved Monte Carlo simulations, the correlated photon and electron (positron) polarization effects in  multiphoton Breit-Wheeler pair production process during the interaction  of  an ultrarelativistic electron  beam  with  a  counterpropagating  elliptically polarized laser pulse.
% by using a fully polarization resolved Monte Carlo method.
We showed that the polarization of $e^-e^+$ pairs is degraded by $35\%$, when the polarization of the intermediate photon is resolved, accompanied with an approximately $13\%$ decrease of the pair yield.
%density. Moreover,  energetic positrons in small angle region reverse the polarization direction, which originates from the pair production of hard photons with polarization parallel with electric field.
Moreover, the polarization direction of energetic positrons in small angle region is reversed, which originates from the pair production of hard photons with polarization parallel with electric field.
\end{abstract}

\maketitle

\section{Introduction}\label{sec:level1}

Polarized positron beams are a powerful tool for exploration of fine structure of matter, in particular, for probing nuclear
%structure
constituents\cite{voutier2014physics}, and testing the validity of the Standard Model\cite{qweak2013first} of particle physics via weak and electromagnetic interactions.
%essential%\cite{shikin1998phonon, voutier2014physics} for reach new  frontier in exploration of nuclear and testing the Standard Model\cite{qweak2013first} of particle physics, which can be obtained via the weak and the electromagnetic interactions.
The natural decay of some radio-isotopes generates polarized positrons with polarization up to $40\%$\cite{mecking2003cebaf}, but the flux is too low for further acceleration and applications. Positrons can also be polarized
%be achieved by self-polarization
in a storage ring due to spin-flips at photon emissions (Sokolov-Ternov effect)\cite{sokolov1964polarization, ternov1995synchrotron, baier1967radiational, bauier1972radiative, derbenev1973polarization}, which however, is a slow process lasting at least several minutes  and can be realized in a large scale storage ring facilities.
%requires the construction of a hundreds of meters scale storage ring.
At particle accelerators
% facilities,
polarized $e^-e^+$ pairs are commonly produced by scattering of circularly polarized gamma-photons in a high-Z material
%nuclear field
via Bethe-Heitler process\cite{sakai2003production, omori2006efficient, ji1997deeply}, while the luminosity of positrons is %rather low due to restriction
limited because of constraints on the target thickness\cite{potylitsin1997production}, and  the  required intense flux of sufficiently energetic photons with a high degree of circular polarization is challenging to produce\cite{li2020polarized}.
%The nuclear and high-energy physics communities have shown a growing interest in the availability of producing high current, highly-polarized positron beams.

Recently, the rapid development of petawatt (PW) laser technology\cite{danson2015petawatt, yanovsky2008ultra,Yoon_2021,Vulcan,ELI,XCELS} and laser wakefield acceleration\cite{leemans2014multi,gonsalves2019petawatt} stimulate distinct interest towards
%continuous interest producing
development of polarized positron source via nonlinear Breit-Wheeler (NBW) process\cite{ritus1985quantum,wan2020ultrarelativistic,chen2019polarized,li2020production,Seipt_2020,Ilderton_2020,Dinu_2020,Torgrimsson_2021}. The positrons created in a strong laser field can be polarized due to the energetically preferred orientation of the positron spin along the local magnetic field. However, the challenge is that in a symmetric laser field (e.g. in a monochromatic laser field) the polarization of positrons created in different laser half-cycles oscillates
%with the laser cycle
following the laser magnetic field and averages out to zero for the total beam.
%Since the positrons generated in strong laser fields are probable to spin parallel with local magnetic field,  the polarization of positrons oscillates with laser cycles and averages out to zero in a symmetric laser field.
Thus, to  achieve net polarization of the created positrons, it is necessary to use an asymmetric laser field. For instance, recently a two-color laser field has been proposed to exploit for ultrafast generation of highly polarized electron\cite{seipt2019ultrafast,song2019spin} and positron beams\cite{chen2019polarized}. Another efficient way for laser driven generation of
%obtaining
polarized electrons (positrons) has been also demonstrated\citep{li2019ultrarelativistic,wan2020ultrarelativistic}, employing the spin-dependent radiation reaction in an elliptically polarized laser pulse  to split the electron (positron) beam  into two oppositely transversely polarized parts. Laser driven positron generation schemes
%driven by ultrastrong laser fields
provide a promising avenue for high current, highly-polarized positron sources.

Usually the gamma-photon which creates a pair in NBW process is generated due to the Compton scattering of incoming electron beam off a counterpropagating laser field. In most of studies, the gamma-photon have been assumed unpolarized and the NBW probability has been averaged over the photon polarization \cite{seipt2019ultrafast,song2019spin,wan2020ultrarelativistic,chen2019polarized}. In reality the intermediate gamma-photon is partially polarized, which has consequences for further pair production process\cite{king2013photon,wan2020high,li2020production}.
In particular,  the decrease of the pair density  with the inclusion of photon polarization has been shown in analytical QED calculations\cite{king2013photon} averaged by the lepton spins, which is confirmed by more accurate spin resolved Monte Carlo simulations\cite{wan2020high}. Moreover, highly polarized gamma-photons can be obtained with polarized seed electrons\citep{li2020polarized}, which in further NBW process may create highly polarized positrons, as it is shown in the Monte Carlo simulation\cite{li2020production} with the use of a simplified pair production probability summed up over final spin states of either electron or positron.
Therefore, including photon polarization in the description of NBW process is mandatory for a reliable prediction of parameters of the laser driven polarized positron source.  The study of fully polarization resolved NBW is also of pure theoretical interest, providing insight on correlations of electrons, positrons and photons polarization in strong-field pair production processes.
%Strong laser driven positron generation schemes  significantly enhances the yield of positrons,  providing promising availability of producing high current, highly-polarized positron beams. However, most studies refer to unpolarized photons where the photon polarization is averaged over\cite{wan2020ultrarelativistic,chen2019polarized}.  Including photon polarization in the description of NBW process is complementary for a reliable proposal of laser driven polarized positron source and is also of pure theoretical interest in the perspective of understanding correlations of electrons, positrons and photons polarization in pair production process.

Analytical description of strong-field QED processes is possibly only in a case of a plane wave laser field \cite{gol1964intensity,nikishov1964quantum,ritus1985quantum,Kotkin_2003,Ivanov_2004,Ivanov_2005}. For QED processes in more realistic scenarios including focused laser fields and laser-plasma interaction, a Monte Carlo method has been developed\cite{ridgers2014modelling, elkina2011qed, green2015simla}, which is based on the local constant field approximation (LCFA) \cite{katkov1998electromagnetic,dinu2016quantum, di2018implementing, ilderton2019extended, podszus2019high, ilderton2019note, di2019improved}, applicable for intense laser-plasma\cite{ridgers2014modelling, elkina2011qed, green2015simla, gonoskov2015extended}/ultra-relativistic electrons interactions\cite{li2020production,chen2019polarized,li2020polarized,wan2020high}.
Recently, the QED Monte Carlo method has been generalized to include spin of involved leptons \cite{li2019ultrarelativistic,chen2019polarized} and polarization of emitted or absorbed photons \cite{li2020polarized,li2020production,wan2020high,king2020nonlinear}.  A numerical approach suitable for treating polarization effects beyond LCFA and plane wave approximations at intermediate laser intensities has been developed\cite{Wistisen_2019,wistisen2020numerical}  for strong-field pair production process within the semiclassical formalism of Baier-Katkov.
%including arbitrary electron spin and photon polarization  within the semiclassical formalism of Baier-Katkov.

%The polarization of emitted photons in nonlinear Compton scattering \cite{gol1964intensity, nikishov1964quantum, bula1996observation} has been investigated with a Monte Carlo method\cite{ridgers2014modelling, elkina2011qed, green2015simla} including photon polarization in radiative process\cite{li2020polarized,king2020nonlinear}.  In NBW process highly polarized gamma photons can be obtained with polarized seed electrons\citep{li2020polarized}. Later,  the photon polarization effects on pair density have been investigated via QED calculation\cite{king2013photon} and Monte Carlo simulation\cite{wan2020high}, revealing a decrease of pair density with the inclusion of photon polarization. Recently, the polarization resolved Monte Carol method is further developed to include the  photon polarization in NBW process but with a simplified pair production probability\cite{li2020production}, where a summation is carried out over final spin states of either electron or positron.Moreover, Wistisen has performed a semiclassical calculation of pair production for arbitrary electron spin and photon polarization in a laser field and developed a numerical approach suitable for intermediate laser intensity\cite{wistisen2020numerical}.

In this paper, we investigate the interaction of an ultrarelativistic electron beam head-on colliding with an ultraintense laser pulse and  focus on the effects of photon polarization in NBW pair production process. To provide an accurate analysis of the produced pair polarization, we employ fully polarization resolved NBW probabilities, i.e., resolved as in incoming photon polarization as well as in the created electron and positron polarizations. The probabilities are derived
%For this purpose, we derive the photon polarized NBW probability
with Baier-Katkov QED operator method \cite{baier1989quantum, katkov1998electromagnetic} within LCFA.
%\cite{dinu2016quantum, di2018implementing, ilderton2019extended, podszus2019high, ilderton2019note, di2019improved}, applicable for intense laser-plasma\cite{ridgers2014modelling, elkina2011qed, green2015simla, gonoskov2015extended}/ultra-relativistic electrons interactions\cite{li2020production,chen2019polarized,li2020polarized,wan2020high},
and have been included into the recently developed laser-electron beam simulation code\cite{li2020production}. We consider a scheme where the initial electrons are transversely polarized and the laser field is elliptically polarized. With the fully polarization-resolved Monte Carlo method,  we find that the polarization
of the produced positrons is highly dependent on the polarization of parent photons.
%In the case of nonlinear Compton scattering of an elliptically polarized laser pulse,
In particular, the polarization of positrons is reduced by 35$\%$ since the emitted photons is partially polarized along electric field direction, and that the angular distribution of positron polarization exhibits an abnormal twist near small angle region, which originates from pair production of highly polarized photons at the high energy end of the spectrum.

\section{Simulation method}

In this section, we analysis the correlation of  photon and positron/electron polarization based on fully polarization resolved probabilities and briefly elaborate on the Monte Carlo method used for simulation.

\subsection{Photon polarization resolved radiation probability}

Here, we provide probabilities of a polarized photon emission with a polarized electron.
Let us assume that the polarization of the emitted photon is $\vec{e}=a_1\vec{e_1}+a_2\vec{e_2}$, where
\begin{equation}
\vec{e}_{1}=\vec{s}-\left(\vec{n}\vec{s}\right)\vec{s},\quad\vec{e}_{2}=\vec{n}\times\vec{e}_{1}.
\end{equation}
$\vec{n}=\vec{k}/\left|\vec{k}\right|$ and $\vec{s}=\vec{w}/\left|\vec{w}\right|$ are the unit vectors along the photon emission and acceleration directions, respectively. The photon polarization resolved emission probability reads
\begin{align}\nonumber
dW_{r} & =\frac{1}{2}\left(dW_{11}+dW_{22}\right)+\frac{\xi_{1}}{2}\left(dW_{11}-dW_{22}\right)\\\nonumber
 & -i\frac{\xi_{2}}{2}\left(dW_{21}-dW_{12}\right)+\frac{\xi_{3}}{2}\left(dW_{11}-dW_{22}\right)\\
 & =\frac{1}{2}\left(F_{0}+\xi_{1}F_{1}+\xi_{2}F_{2}+\xi_{3}F_{3}\right),
\end{align}
where $\xi_i$ $(i=1,2,3)$ are Stokes parameters with respect to axes ($\vec{e_1}$, $\vec{e_2}$, $\vec{n}$), and
\begin{align}\nonumber
F_0 & =\frac{\alpha}{2\sqrt{3}\pi\gamma^{2}}d\omega\left\{ \left(\frac{\varepsilon^{2}+\varepsilon'^{2}}{\varepsilon\varepsilon'}\textrm{K}_{\frac{2}{3}}\left(z_{q}\right)-\intop_{z_{q}}^{\infty}dx\textrm{K}_{\frac{1}{3}}\left(x\right)\right)\right.\\\nonumber
 & +\left(2\textrm{K}_{\frac{2}{3}}\left(z_{q}\right)-\intop_{z_{q}}^{\infty}dx\textrm{K}_{\frac{1}{3}}\left(x\right)\right)(\vec{\zeta}_{i}\vec{\zeta}_{f})\\\nonumber
 & -\textrm{K}_{\frac{1}{3}}\left(z_{q}\right)\left(\frac{\omega}{\varepsilon}(\vec{\zeta}_{i}\vec{b})+\frac{\omega}{\varepsilon'}(\vec{\zeta}_{f}\vec{b})\right)\\
 & \left.+\frac{\omega^{2}}{\varepsilon'\varepsilon}\left(\textrm{K}_{\frac{2}{3}}\left(z_{q}\right)-\intop_{z_{q}}^{\infty}dx\textrm{K}_{\frac{1}{3}}\left(x\right)\right)\left(\vec{\zeta}_{i}\vec{\hat{v}}\right)\left(\vec{\zeta}_{f}\vec{\hat{v}}\right)\right\},
\end{align}

\begin{align}\nonumber
F_3 & =\frac{\alpha}{2\sqrt{3}\pi\gamma^{2}}d\omega\left\{ \textrm{K}_{\frac{2}{3}}\left(z_{q}\right)+\frac{\varepsilon^{2}+\varepsilon'^{2}}{2\varepsilon'\varepsilon}\textrm{K}_{\frac{2}{3}}\left(z_{q}\right)(\vec{\zeta}_{i}\vec{\zeta}_{f})\right.\\\nonumber
 & -\left[\frac{\omega}{\varepsilon'}(\vec{\zeta}_{i}\vec{b})+\frac{\omega}{\varepsilon}(\vec{\zeta}_{f}\vec{b})\right]\textrm{K}_{\frac{1}{3}}\left(z_{q}\right)\\\nonumber
 & +\frac{\omega^{2}}{2\varepsilon'\varepsilon}\left(-\textrm{K}_{\frac{2}{3}}\left(z_{q}\right)\left(\vec{\zeta}_{i}\vec{\hat{v}}\right)\left(\vec{\zeta}_{f}\vec{\hat{v}}\right)\right.\\
 & \left.\left.+\intop_{z_{q}}^{\infty}dx\textrm{K}_{\frac{1}{3}}\left(x\right)\left(\left[\vec{\zeta}_{i}\vec{b}\right]\left[\vec{\zeta}_{f}\vec{b}\right]-\left(\vec{\zeta}_{i}\vec{s}\right)\left(\vec{\zeta}_{f}\vec{s}\right)\right)\right)\right\},
\end{align}

\begin{align}\nonumber
F_1 & =\frac{\alpha}{2\sqrt{3}\pi\gamma^{2}}d\omega\left\{ \frac{\varepsilon^{2}-\varepsilon'^{2}}{2\varepsilon'\varepsilon}\textrm{K}_{\frac{2}{3}}\left(z_{q}\right)\left(\vec{\hat{v}}\left[\vec{\zeta}_{f}\times\vec{\zeta}_{i}\right]\right)\right.\\\nonumber
 & +\left[\frac{\omega}{\varepsilon'}(\vec{\zeta}_{i}\vec{s})+\frac{\omega}{\varepsilon}(\vec{\zeta}_{f}\vec{s})\right]\textrm{K}_{\frac{1}{3}}\left(z_{q}\right)\\
 & \left.-\frac{\omega^{2}}{2\varepsilon'\varepsilon}\intop_{z_{q}}^{\infty}dx\textrm{K}_{\frac{1}{3}}\left(x\right)\left(\left(\vec{\zeta}_{i}\vec{s}\right)\left(\vec{\zeta}_{f}\vec{b}\right)+\left(\vec{\zeta}_{i}\vec{b}\right)\left(\vec{\zeta}_{f}\vec{s}\right)\right)\right\},
\end{align}

\begin{align}\nonumber
F_2 & =-\frac{\alpha}{2\sqrt{3}\pi\gamma^{2}}d\omega\left\{ \frac{\varepsilon^{2}-\varepsilon'^{2}}{2\varepsilon'\varepsilon}\textrm{K}_{\frac{1}{3}}\left(z_{q}\right)\left(\vec{s}\left[\vec{\zeta}_{f}\times\vec{\zeta}_{i}\right]\right)\right.\\\nonumber
 & +\left[-\frac{\varepsilon^{2}-\varepsilon'^{2}}{\varepsilon'\varepsilon}\textrm{K}_{\frac{2}{3}}\left(z_{q}\right)+\frac{\omega}{\varepsilon}\intop_{z_{q}}^{\infty}dx\textrm{K}_{\frac{1}{3}}\left(x\right)\right](\vec{\zeta}_{i}\vec{\hat{v}})\\\nonumber
 & +\left[-\frac{\varepsilon^{2}-\varepsilon'^{2}}{\varepsilon'\varepsilon}\textrm{K}_{\frac{2}{3}}\left(z_{q}\right)+\frac{\omega}{\varepsilon'}\intop_{z_{q}}^{\infty}dx\textrm{K}_{\frac{1}{3}}\left(x\right)\right](\vec{\zeta}_{f}\vec{\hat{v}})\\
 & \left.+\frac{\omega^{2}}{2\varepsilon'\varepsilon}\textrm{K}_{\frac{1}{3}}\left(z_{q}\right)\left(\left(\vec{\zeta}_{i}\vec{\hat{v}}\right)\left(\vec{\zeta}_{f}\vec{b}\right)+\left(\vec{\zeta}_{i}\vec{b}\right)\left(\vec{\zeta}_{f}\vec{\hat{v}}\right)\right)\right\},
\end{align}
where $\vec{\hat{v}}=\frac{\vec{v}}{v},\vec{b}=\vec{\hat{v}}\times\vec{s},z_{q}=\frac{2}{3}\frac{\omega}{\chi_{e}\varepsilon'}$, $\varepsilon$ and $\varepsilon'$  are the energy of the emitting particle before and after emission, respectively, $\vec{\zeta_i}$ and $\vec{\zeta_f}$ are the spin vectors before and after emission, respectively, $\omega$ is the emitted photon energy.

\begin{figure}
	\setlength{\abovecaptionskip}{-0.2cm}
	\includegraphics[width=0.5\textwidth]{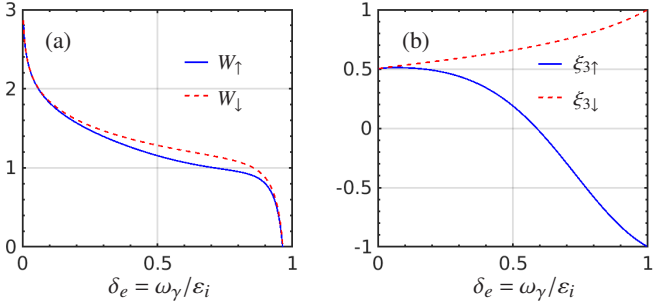}
	\begin{picture}(300,20)
	\put(20,108){(a)}
	\put(155,108){(b)}
	\put(44,15){$\delta_e=\omega_\gamma/\varepsilon_i$}
	\put(182,15){$\delta_e=\omega_\gamma/\varepsilon_i$}
	\put(85,100){\footnotesize$W_{\uparrow}$}
	\put(85,86){\footnotesize$W_{\downarrow}$}
	\put(220,100){\footnotesize$\xi_{3\uparrow}$}
	\put(220,86){\footnotesize$\xi_{3\downarrow}$}
	\end{picture}
	\caption{(a) The photon emission probability log$_{10}W_{i}$ (arb. units), and (b) Stokes parameter $\xi_{3i}$ ($i \in{\uparrow,\downarrow}$) vs emitted photon energy $\delta_e=\omega_\gamma/\varepsilon_i$ for $\chi_e=10$. $i$ denotes the electron spin before the emission with respect to the magnetic field direction.}
		\label{Fig. 1}
\end{figure}

The emission probability and the polarization of the emitted photon both depend on the initial electron spin $\vec{\zeta_i}$. For instance, the emission probability is larger for the spin-down electron, with respect to magnetic field direction in the rest frame of the electron, than for the spin-up electrons, as shown in  Fig. \ref{Fig. 1}(a). The dependence of polarization of the photon on $\vec{\zeta_i}$ is more remarkable, as shown in  Fig. \ref{Fig. 1} (b).
For low energy region, the Stokes parameter $\xi_3\sim0.5$ regardless of the initial electron spin. However, with the increase of the emitted photons energy, $\xi_3$ increases up to $\xi_3=1$ for the spin-up electron, while decreases to $\xi_3=-1$ for the opposite case.

\subsection{Photon polarization resolved pair production probabilities}

Here, we provide probability of a polarized electron-positron pair production with a polarized photon.

\begin{comment}
\begin{eqnarray}\nonumber
dW_{p}&=&\frac{1}{2}\frac{\alpha}{\left(2\pi\right)^{2}}\frac{dp^{3}}{\omega}\int d\tau\underset{\lambda}{\sum}R_{p}^{*}\left(t-\frac{\tau}{2}\right)R_{p}\left(t+\frac{\tau}{2}\right)\\
&&\exp\left\{ i\frac{\varepsilon}{\varepsilon_{+}}\left[kx\left(t+\frac{\tau}{2}\right)-kx\left(t-\frac{\tau}{2}\right)\right]\right\}
\end{eqnarray}
\end{comment}
The polarization of the photon is defined as follows:
%$\vec{e}=a_1\vec{e_1}+a_2\vec{e_2}$, where
\begin{eqnarray}
\vec{e}&=&a_1\vec{e_1}+a_2\vec{e_2}\\
\vec{e}_{1}&=&\frac{\vec{E}-\vec{n}\left(\vec{n}\vec{E}\right)+\vec{n}\times\vec{B}}{\left|\vec{E}-\vec{n}\left(\vec{n}\vec{E}\right)+\vec{n}\times\vec{B}\right|},\\ \vec{e}_{2}&=&\vec{n}\times\vec{e}_{1}, \vec{n}=\frac{\vec{k}}{\left|\vec{k}\right|}.
\end{eqnarray}
The pair production rate of the polarized photon takes the form
\begin{align}\nonumber
dW_{p} & =\frac{1}{2}\left(dW^{(11)}+dW^{(22)}\right)+\frac{\xi_{1}}{2}\left(dW^{(11)}-dW^{(22)}\right)\\\nonumber
 & -i\frac{\xi_{2}}{2}\left(dW^{(21)}-dW^{(12)}\right)+\frac{\xi_{3}}{2}\left(dW^{(11)}-dW^{(22)}\right)\\
 & =\frac{1}{2}\left(G_{0}+\xi_{1}G_{1}+\xi_{2}G_{2}+\xi_{3}G_{3}\right),
\end{align}
where $\xi_i=\frac{G_i}{G_0}, i=1,2,3$ are the Stokes parameters.
\begin{align}\nonumber
G_0&  =\frac{\alpha m^{2}d\varepsilon}{2\sqrt{3}\pi\omega^{2}}\Bigg\{\left\{ \intop_{z_{p}}^{\infty}dx\textrm{K}_{\frac{1}{3}}\left(x\right)+\frac{\varepsilon_{+}^{2}+\varepsilon^{2}}{\varepsilon_{+}\varepsilon}\textrm{K}_{\frac{2}{3}}\left(z_{p}\right)\right\} \\\nonumber
 & +\left\{ \intop_{z_{p}}^{\infty}dx\textrm{K}_{\frac{1}{3}}\left(x\right)-2\textrm{K}_{\frac{2}{3}}\left(z_{p}\right)\right\} \left(\vec{\zeta}_{-}\vec{\zeta}_{+}\right)\\\nonumber
 & +\left[\frac{\omega}{\varepsilon_{+}}\left(\vec{\zeta}_{+}\vec{b}\right)-\frac{\omega}{\varepsilon}\left(\vec{\zeta}_{-}\vec{b}\right)\right]\textrm{K}_{\frac{1}{3}}\left(z_{p}\right)\\
 & +\left\{ \frac{\varepsilon_{+}^{2}+\varepsilon^{2}}{\varepsilon\varepsilon_{+}}\intop_{z_{p}}^{\infty}dx\textrm{K}_{\frac{1}{3}}\left(x\right)-\frac{\left(\varepsilon_{+}-\varepsilon\right)^{2}}{\varepsilon\varepsilon_{+}}\textrm{K}_{\frac{2}{3}}\left(z_{p}\right)\right\} \left(\vec{\zeta}_{-}\vec{\hat{v}}\right)\left(\vec{\zeta}_{+}\vec{\hat{v}}\right)\Bigg\}
\end{align}

\begin{align}\nonumber
G_3 & =\frac{\alpha m^{2}d\varepsilon}{2\sqrt{3}\pi\omega^{2}}\Bigg\{-\textrm{K}_{\frac{2}{3}}\left(z_{p}\right)+\frac{\varepsilon_{+}^{2}+\varepsilon^{2}}{2\varepsilon_{+}\varepsilon}\textrm{K}_{\frac{2}{3}}\left(z_{p}\right)\left(\vec{\zeta}_{-}\vec{\zeta}_{+}\right)\\\nonumber
 & +\left[-\frac{\omega}{\varepsilon}\left(\vec{\zeta}_{+}\vec{b}\right)+\frac{\omega}{\varepsilon_{+}}\left(\vec{\zeta}_{-}\vec{b}\right)\right]\textrm{K}_{\frac{1}{3}}\left(z_{p}\right)\\\nonumber
 & -\frac{\left(\varepsilon_{+}-\varepsilon\right)^{2}}{2\varepsilon_{+}\varepsilon}\textrm{K}_{\frac{2}{3}}\left(z_{p}\right)\left(\vec{\zeta}_{-}\vec{\hat{v}}\right)\left(\vec{\zeta}_{+}\vec{\hat{v}}\right)\\
 & +\frac{\omega^{2}}{2\varepsilon_{+}\varepsilon}\intop_{z_{p}}^{\infty}dx\textrm{K}_{\frac{1}{3}}\left(x\right)\left[\left(\vec{\zeta}_{-}\vec{b}\right)\left(\vec{\zeta}_{+}\vec{b}\right)-\left(\vec{\zeta}_{-}\vec{s}\right)\left(\vec{\zeta}_{+}\vec{s}\right)\right]\Bigg\}
\end{align}

\begin{align}\nonumber
G_1 & =\frac{\alpha m^{2}d\varepsilon}{2\sqrt{3}\pi\omega^{2}}\Bigg\{-\frac{\varepsilon_{+}^{2}-\varepsilon^{2}}{2\varepsilon_{+}\varepsilon}\textrm{K}_{\frac{2}{3}}\left(z_{p}\right)\vec{\hat{v}}(\vec{\zeta}_{+}\times\vec{\zeta}_{-})\\\nonumber
 & +\left[\frac{\omega}{\varepsilon}\left(\vec{\zeta}_{+}\vec{s}\right)-\frac{\omega}{\varepsilon_{+}}\left(\vec{\zeta}_{-}\vec{s}\right)\right]\textrm{K}_{\frac{1}{3}}\left(z_{p}\right)\\
 & -\frac{\omega^{2}}{2\varepsilon_{+}\varepsilon}\intop_{z_{p}}^{\infty}dx\textrm{K}_{\frac{1}{3}}\left(x\right)\left\{ \left(\vec{\zeta}_{-}\vec{b}\right)\left(\vec{\zeta}_{+}\vec{s}\right)+\left(\vec{\zeta}_{-}\vec{s}\right)\left(\vec{\zeta}_{+}\vec{b}\right)\right\} \Bigg\}
\end{align}

\begin{align}\nonumber
G_2 & =\frac{\alpha m^{2}d\varepsilon}{2\sqrt{3}\pi\omega^{2}}\Bigg\{-\frac{\omega^{2}}{2\varepsilon_{+}\varepsilon}\textrm{K}_{\frac{1}{3}}\left(z_{p}\right)\left[\vec{s}(\vec{\zeta}_{-}\times\vec{\zeta}_{+})\right]\\\nonumber
 & +\left(\frac{\omega}{\varepsilon_{+}}\intop_{z_{p}}^{\infty}dx\textrm{K}_{\frac{1}{3}}\left(x\right)+\frac{\varepsilon_{+}^{2}-\varepsilon^{2}}{\varepsilon_{+}\varepsilon}\textrm{K}_{\frac{2}{3}}\left(z_{p}\right)\right)\left(\vec{\zeta}_{+}\vec{\hat{v}}\right)\\\nonumber
 & +\left(\frac{\omega}{\varepsilon}\intop_{z_{p}}^{\infty}dx\textrm{K}_{\frac{1}{3}}\left(x\right)-\frac{\varepsilon_{+}^{2}-\varepsilon^{2}}{\varepsilon_{+}\varepsilon}\textrm{K}_{\frac{2}{3}}\left(z_{p}\right)\right)\left(\vec{\zeta}_{-}\vec{\hat{v}}\right)\\
 & -\frac{\varepsilon_{+}^{2}-\varepsilon^{2}}{2\varepsilon_{+}\varepsilon}\textrm{K}_{\frac{1}{3}}\left(z_{p}\right)\left[\left(\vec{\zeta}_{-}\vec{\hat{v}}\right)\left(\vec{\zeta}_{+}\vec{b}\right)+\left(\vec{\zeta}_{-}\vec{b}\right)\left(\vec{\zeta}_{+}\vec{\hat{v}}\right)\right]\Bigg\},
\end{align}
where $\vec{\hat{v}}=\frac{\vec{v}}{v}$ is the velocity direction of the produced pairs, which fulfills $\vec{\hat{v}}_-\approx\vec{\hat{v}}_+\approx\vec{n}$ in the relativistic case, $\vec{s}=\frac{\vec{w}}{|\vec{w}|}$ is the acceleration direction of the produced particles, $\vec{b}=\vec{\hat{v}}\times\vec{s}$ the magnetic field direction in the rest frame of the electron/positron, the parameter $z_{p}=\frac{2\omega}{3\chi_{\gamma}\varepsilon\varepsilon_{+}}$, where $\omega$, $\varepsilon$ and $\varepsilon_+$ are the energy of the parent photon, the produced electron and positron, respectively, the quantum strong-field parameter $\chi_\gamma=\frac{|e|\sqrt{(F_\mu v k^v)^2}}{m^3}$, with the four-vector $k$ of the photon momentum. Averaging over the photon polarization yields the spin-resolved pair production probability\cite{blackburn2017scaling, olugh2019pair, katkov1998electromagnetic}: $dW_p(0)=\frac{1}{2}G_0$.

Summing up the final spin state of electron,  one can obtain the polarization of the  positron depending on the photon polarization:
\begin{equation}\label{PP1}
\vec{\zeta_{+}}^{f,\xi}=\frac{\xi_{1}f_{3}\frac{\omega}{\varepsilon}\vec{s}+\xi_{2}\vec{\hat{v}}\left(\frac{\omega}{\varepsilon_{+}}f_{1}+\frac{\varepsilon_{+}^{2}-\varepsilon^{2}}{\varepsilon\varepsilon_{+}}f_{2}\right)+\left(\frac{\omega}{\varepsilon_{+}}-\xi_{3}\frac{\omega}{\varepsilon}\right)\vec{b}f_{3}}{f_{1}+\frac{\varepsilon^{2}+\varepsilon_{+}^{2}}{\varepsilon\varepsilon_{+}}f_{2}-\xi_{3}f_{2}}.
\end{equation}
In the case of an unpolarized photon:
\begin{equation}\label{PP2}
\vec{\zeta_{+}}^{f,0}=\frac{\frac{\omega}{\varepsilon_{+}}\vec{b}f_{3}}{f_{1}+\frac{\varepsilon^{2}+\varepsilon_{+}^{2}}{\varepsilon\varepsilon_{+}}f_{2}},
\end{equation}
where $f_1=\intop_{z_{p}}^{\infty}dx\textrm{K}_{\frac{1}{3}}\left(x\right)$, $f_2=\textrm{K}_{\frac{2}{3}}\left(z_{p}\right)$, $f_3=\textrm{K}_{\frac{1}{3}}\left(z_{p}\right)$.  Eqs (\ref{PP1}) and (\ref{PP2}) show that  the photon polarization  has significant effects on positron polarization $\vec{\zeta_{+}}^{f}$. The longitudinal polarization of positrons is completely missing if the photon polarization is averaged, while the transverse polarization either increase or decrease determined by $\xi_1$ and $\xi_3$.

\begin{figure}
	\includegraphics[width=0.48\textwidth]{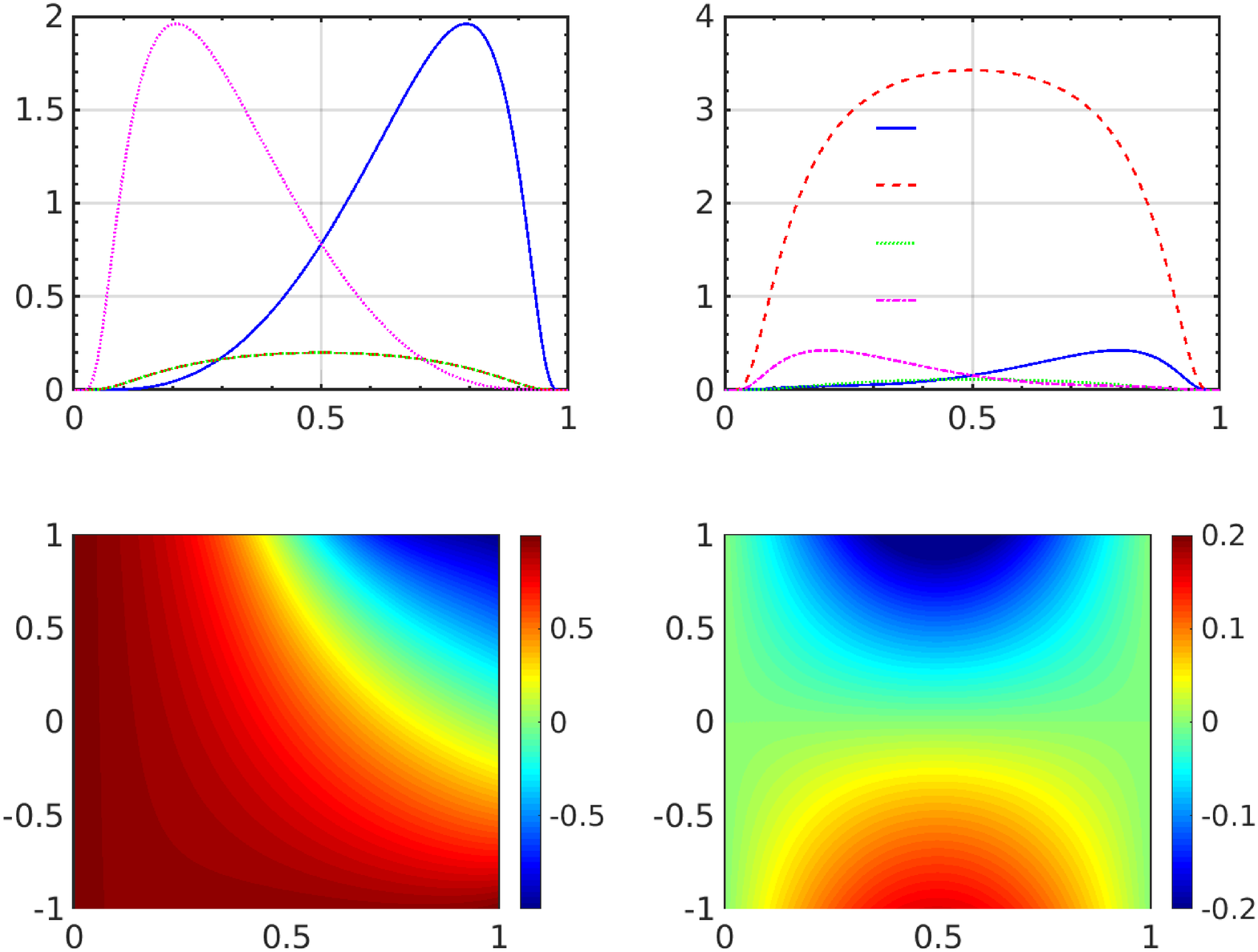}
	\begin{picture}(300,25)
	\put(18,197){(a)}
	\put(143,197){(b)}
	\put(18,96){(c)}
	\put(143,96){(d)}
	\put(55,15){$\delta_+$}
	\put(180,15){$\delta_+$}
	\put(60,116){$\delta_+$}
	\put(186,116){$\delta_+$}
	%\put(49,210){$\xi_3=0.55$}
	%\put(175,210){$\xi_3=1.0$}
	%\put(49,109){$\xi_3=-1.0$}
    \put(180,185){\footnotesize$dW_{\downarrow\downarrow}$}
    \put(180,174){\footnotesize$dW_{\downarrow\uparrow}$}
    \put(180,163){\footnotesize$dW_{\uparrow\downarrow}$}
    \put(180,150){\footnotesize$dW_{\uparrow\uparrow}$}
    \put(-7,67){\rotatebox{90}{$\xi_3$}}
    \put(-7,140){\rotatebox{90}{$dW_p$(arb. units)}}
    \put(124,67){\rotatebox{90}{$\xi_3$}}
    \put(120,140){\rotatebox{90}{$dW_p$(arb. units)}}
	\end{picture}
	\vskip -1em
	\includegraphics[width=0.4\textwidth]{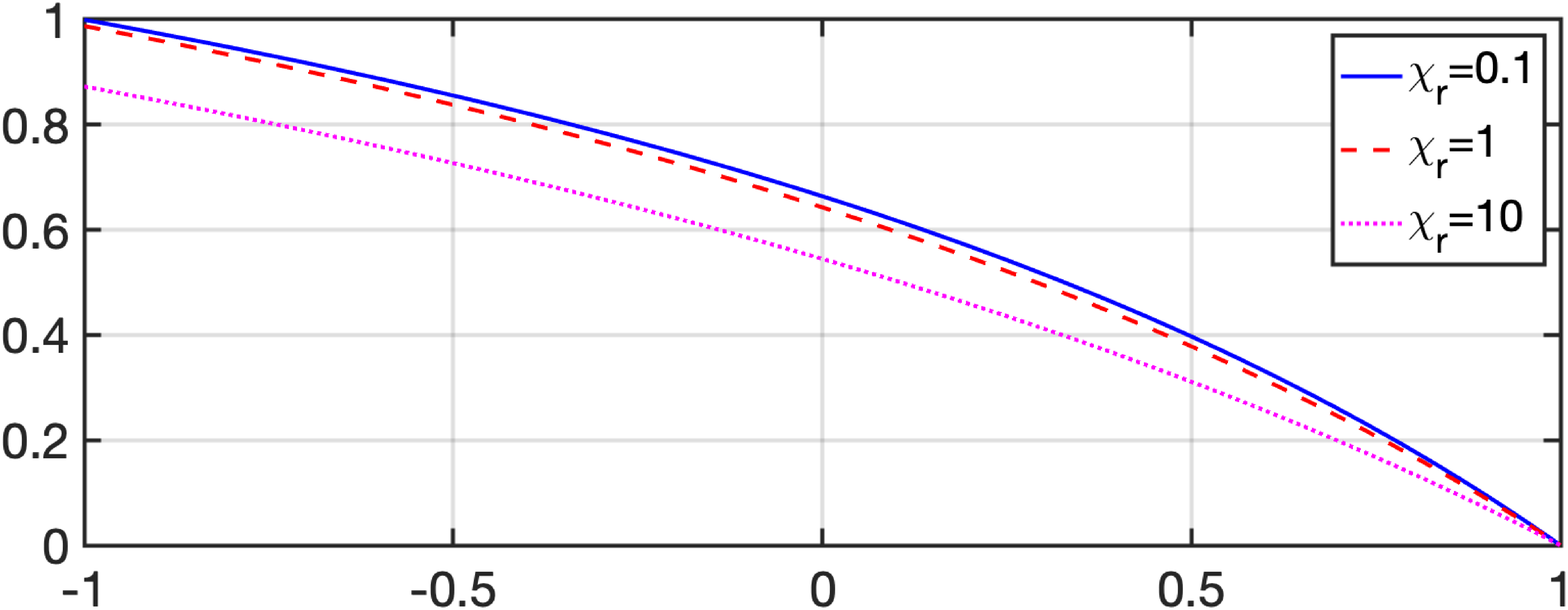}
    \begin{picture}(300,25)
    \put(125,20){$\xi_3$}
    \put(10,70){\rotatebox{90}{$\zeta_+$}}
    \put(170,90){(e)}
    \end{picture}
    	\vspace{-3.5em}
	\caption{The pair production probability $dW_{\zeta_-\zeta_+}$ for the polarized photon with (a) $\xi_1=\xi_2=0$, $\xi_3=1$, and (b) $\xi_3=-1$;  (c) Positron polarization $\zeta_+=\sum_{\zeta_-}\frac{dW_{\zeta_-\uparrow}-dW_{\zeta_-\downarrow}}{dW_{\zeta_-\uparrow}+dW_{\zeta_-\downarrow}}$ vs  $\xi_3$ and $\delta_+$; (d) Difference of the photon polarization resolved and averaged pair production probabilities $\frac{dW_p(\xi)-dW_p(0)}{dW_p(\xi)+dW_p(0)}$ vs  $\xi_3$ and $\delta_+$; $\chi_\gamma=3$ for (a),(b) and (c); (e) $\overline{\zeta}_+=\sum_i\zeta_{+}(\delta_i)\cdot \frac{dW_p(\delta_i)}{\sum_idW_p(\delta_i)}$ versus $\xi_3$ for $\chi_\gamma=0.1$ (blue-solid), $\chi_\gamma=1$ (red-dashed) and $\chi_\gamma=10$ (magenta-dotted).}
	\label{Fig. 2}
\end{figure}

The correlation of the electron and positron polarizations in the pair production is analysed in Fig.~\ref{Fig. 2}.
In the case  $\xi_1=\xi_2=0$ and $\xi_3>0$,  the probabilities of  $e^+e^-$ co-polarization are higher than counter-polarization with respect to magnetic field direction, i.e. $dW_{\uparrow\uparrow}, dW_{\downarrow\downarrow}>dW_{\downarrow\uparrow},dW_{\uparrow\downarrow}$, as shown in Fig. \ref{Fig. 2}(a).
% This difference in probability grows with the increase of $\xi_3$. When $\xi_3=1$,  $dW_{\uparrow\uparrow}, dW_{\downarrow\downarrow}$ prevail over $dW_{\downarrow\uparrow},dW_{\uparrow\downarrow}$, as shown in Fig. \ref{Fig. 0} (b).
On the other hand, when $\xi_3<0$, % the amplitude of the parent photon being polarized along $\vec{e}_2$ is larger than that along $\vec{e}_1$. In this case,
the probability of producing an electron with spin down and positron spin up dominates, i.e. $dW_{\downarrow\uparrow}>dW_{\downarrow\downarrow},dW_{\uparrow\uparrow},dW_{\uparrow\downarrow}$, as shown in Fig. \ref{Fig. 2}(b).

After integration over the electron spin, one obtain the dependence of positron spin on the positron energy and the polarization of the parent photon, as shown in Fig. \ref{Fig. 2}(c). For $\xi_3<0$, positron polarization degree decreases gradually with the increase of positron energy.  The domination of $dW_{\downarrow\uparrow}$ results in a high polarization degree of positrons with spin up through the whole spectrum. While for a photon with $\xi_3>0$, the polarization of the produced positron decreases dramatically to a negative value with the increase of energy, resulting in a smaller averaged polarization compared with the case of $\xi_3=0$. Especially, when $\xi_3\sim1$ shown in Fig. \ref{Fig. 2} (a), the probability $\sum_{\zeta_-}dW_{\zeta_-\uparrow}\approx \sum_{\zeta_-}dW_{\zeta_-\downarrow}$. The polarization of positrons in high ($\delta_+>0$) and low energy ($\delta_+<0$) regions cancel each other out,
producing unpolarized positrons after energy integration. Therefore, if the parent photon is polarized along laser polarization direction, i.e. $\xi_3=1$, the produced pairs are unpolarized.
While, if the polarization of the parent photon is orthogonal to laser polarization, i.e. $\xi_3=-1$, the produced pairs have a high degree of polarization with positrons spin up and electrons spin down. 
After integration over the positron energy,  one obtain the  relation between positron polarization and  polarization of its parent photon, as shown in Fig. \ref{Fig. 2} (e). The polarization of positron decreases monotonously with the increase of $\xi_3$, which provides a way of estimating the polarization of the intermediate photons during nonlinear Compton scattering. For instance,  if the polarization of positrons is measured to be 37$\%$ at $\chi_\gamma=1$, the polarization of the intermediate photons is  around $\xi_3=0.5$.

The photon polarization not only affect the polarization of produced pairs, but  also the pairs density, as shown in Fig. \ref{Fig. 2} (d). When $\xi_3>0$,  the pair production probability is smaller than the case where photon polarization is unresolved, i.e. $\xi_3=0$. On the contrary, the photons with $\xi_3<0$ yield more pairs than the unpolarized photons.

%\begin{widetext}
\begin{figure*}
	\setlength{\abovecaptionskip}{-0.2cm}
	\includegraphics[width=0.8\textwidth]{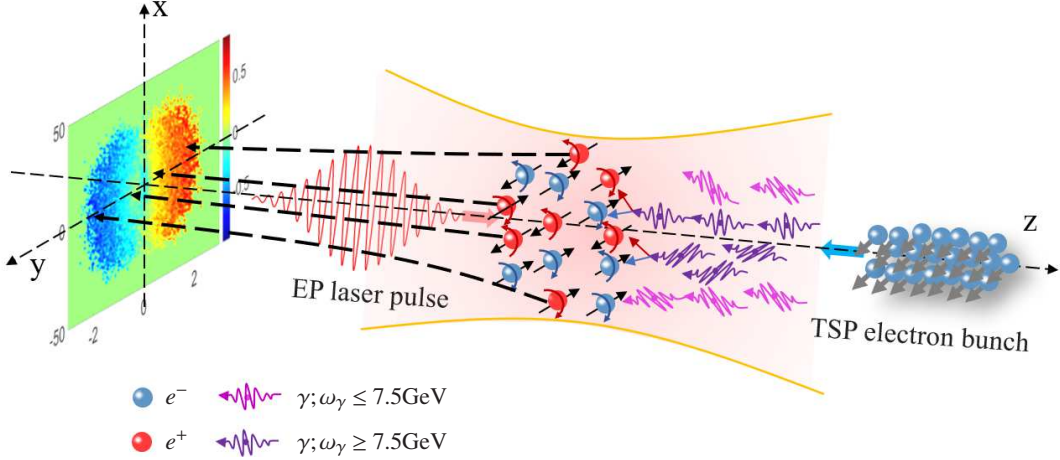}
	\begin{picture}(300,20)
	\put(10,41){$e^-$}
	\put(10,24){$e^+$}
	\put(60,41){$\gamma;\omega_\gamma\leq 7.5$GeV}
	\put(60,24){$\gamma;\omega_\gamma\geq 7.5$GeV}
	\end{picture}
	\caption{Schemes of producing polarized positrons via nonlinear Compton scattering of an initially transversely polarized electrons off a strong elliptically polarized laser pulse. The energetic gamma-photons in $\theta^\gamma<0$ region  have high polarizations up to $\xi_3=1$, resulting in a reduction even reverse of polarization of positrons in small angle region.  }
	\label{Fig. s}
\end{figure*}
%\end{widetext}

\subsection{Stochastic algorithm}

To simulate the nonlinear Compton scattering of a strong laser pulse at an ultrarelativistic electrons beam,  we modified the three-dimensional  Monte Carlo method \cite{li2020production} by employing the fully polarization resolved pair production probability given in Sec.~II.B.  The developed Monte-Carlo method includes the correlation of electron and positron spin, providing complete description of polarization effects in strong-field QED processes. In each simulation step, one  calculates the total emission rate to determine the occurrence of a photon emission, and  the pair production rate to determine the pair production event,  using the common QED Monte Carlo stochastic algorithm\cite{ridgers2014modelling, elkina2011qed, green2015simla}.
%in laser-plasma community.
The spins of electron/positron after emission and creation are determined by the polarization-resolved emission probability of Sec.~II.A and pair production probability of Sec.~II.B, respectively,  according to the stochastic algorithm. The electron/positron spin instantaneously collapses into one of its basis states defined with respect to the instantaneous spin quantization axis (SQA), which is chosen according to the properties of the scattering process. Moreover, since the probability of no emission is also polarization resolved and asymmetric  along arbitrary SQA, it is necessary to include the spin variation between emissions induced by radiative polarization, besides spin precession governed by the Thomas-Bargmann-Michel-Telegdi equation\cite{thomas1926motion,thomas1927kinematics,bargmann1959precession}; for more details, see\cite{li2020production}. The polarization of the emitted photon is determined with a similar stochastic procedure, which also has been used in laser-plasma simulation codes\cite{xue2020generation}.

\section{simulation result}

Recently, various schemes  have been proposed to produce transversely polarized positrons  via strong lasers, however, neglecting the polarization of photon. Here, we proceed to investigate the photon polarization effects on transverse polarization of positrons with the fully polarization resolved Monte-Carlo method, and compare the  result  with that obtained by the unpolarized photon model.
%polarization free model.
A PW laser with intensity $\xi_0=\frac{|e|E_0}{m\omega}=100$ ($I=10^{22}$W/cm$^2$) counterpropagates with a relativistic electron beam with an energy of $\varepsilon_0=10$ GeV, the setup is shown in Fig.~\ref{Fig. s}. The wavelength of the laser is $\lambda_0=1\,\mu$m, the beam waist size $w=5\lambda_0$, the pulse duration $\tau_p=8T$ and the ellipticity $\epsilon=0.03$. The electron beam consists of $N_e=6\times 10^6$ electrons, with the beam length $L_e=5\lambda_0$, the beam radius $r_e=\lambda_0$, the energy divergence $\Delta \varepsilon=0.06$, the angular divergences $\Delta \theta=0.3$ mrad and $\Delta \phi=1$ mrad. The initial  electron beam is fully polarized along $y$-direction.

\begin{figure}
	\includegraphics[width=0.5\textwidth]{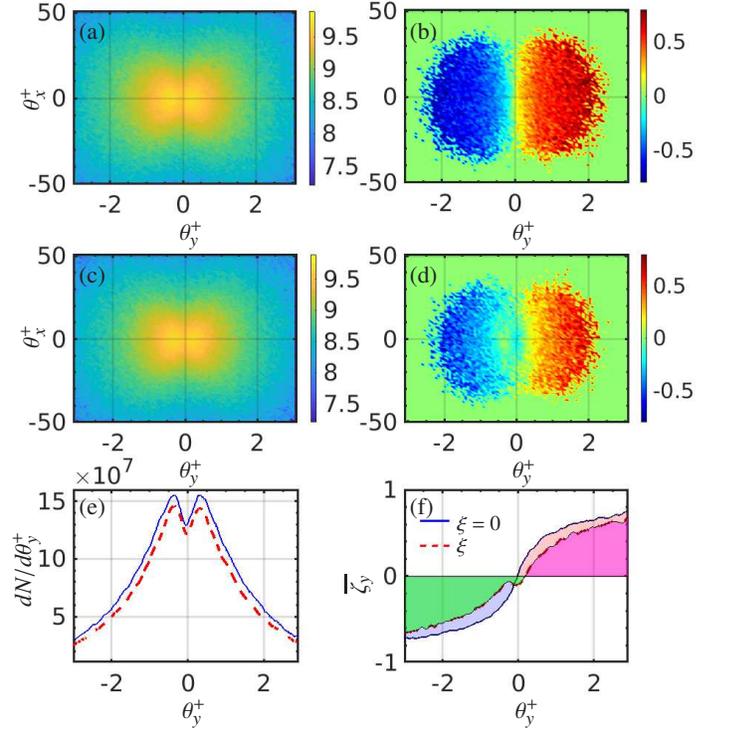}
	\begin{picture}(300,20)
	\put(20,270){(a)}
	\put(145,270){(b)}
	\put(20,178){(c)}
	\put(145,178){(d)}
	\put(20,90){(e)}
	\put(145,90){(f)}
	\put(59,13){$\theta^+_y$}
	\put(185,13){$\theta^+_y$}
	\put(58,104){$\theta^+_y$}
	\put(184,104){$\theta^+_y$}
    \put(58,194){$\theta^+_y$}
	\put(184,194){$\theta^+_y$}
	\put(-2,153){\rotatebox{90}{$\theta^+_x$}}
    \put(-2,243){\rotatebox{90}{$\theta^+_x$}}
    \put(-5,55){\rotatebox{90}{$dN/d\theta^+_y$}}
    \put(120,60){\rotatebox{90}{$\overline{\zeta_y}$}}
    \put(163,84){\footnotesize $\xi=0$}
    \put(163,76){\footnotesize $\xi$}
	\end{picture}
	\caption{Polarized positrons density distribution d$^2N^+/$d$\theta^+_x$d$\theta^+_y$ (rad$^{-2}$)  (left column) and averaged polarization degree of $y$ component $\overline{\zeta_y}$ (right column) versus $\theta^+_x=p_x/p_z$ (rad) and $\theta^+_y=p_y/p_z$ (rad) for photon polarization unresolved (top row) and resolved (middle row) pair production. (bottom row) Polarized positrons density distribution d$N/$d$\theta_y$ (rad$^{-1}$) versus $\theta_y$ (rad) for unresolved (solid line) and resolved photon polarization (dotted line); Averaged polarization degree $\overline{\zeta_y}$ versus $\theta_y$ (rad) for unresolved (dashed line) and resolved photon polarization (dash-dot line). }
	\label{Fig. 3}
\end{figure}

The impact of the photon polarization  on the pair production is elucidated in Fig.~\ref{Fig. 3}.
The positron density decreases when photon polarization is resolved in both photon emission and pair production processes, as shown in Fig.~\ref{Fig. 3}(a),(c),(e). The difference in positron density of using polarization resolved or unresolved treatment is approximately $(N(\xi)-N(0))/N(0)\approx12.6\%$. More importantly, the $y$-component of polarization $\overline{\zeta_y}$ decreases dramatically at small angles, even showing reversal of polarization direction, see Fig. \ref{Fig. 3}(d). As a consequence, the symmetric angular distribution of $\overline{\zeta_y}$ near $\theta_y=0$ is distorted when the intermediate photon polarization is considered, as shown in Fig. \ref{Fig. 3}(f).
The angular distribution of $\overline{\zeta_y}$ oscillates around small angle region, instead of a monotone increase as in the photon polarization averaged case.  The average polarization of positron beam decreases by $[\overline{\zeta_y}(\xi)-\overline{\zeta_y}(0)]/\overline{\zeta_y}(0)\approx35\%$.

\begin{figure}
	\setlength{\abovecaptionskip}{-0.2cm}
	\includegraphics[width=0.5\textwidth]{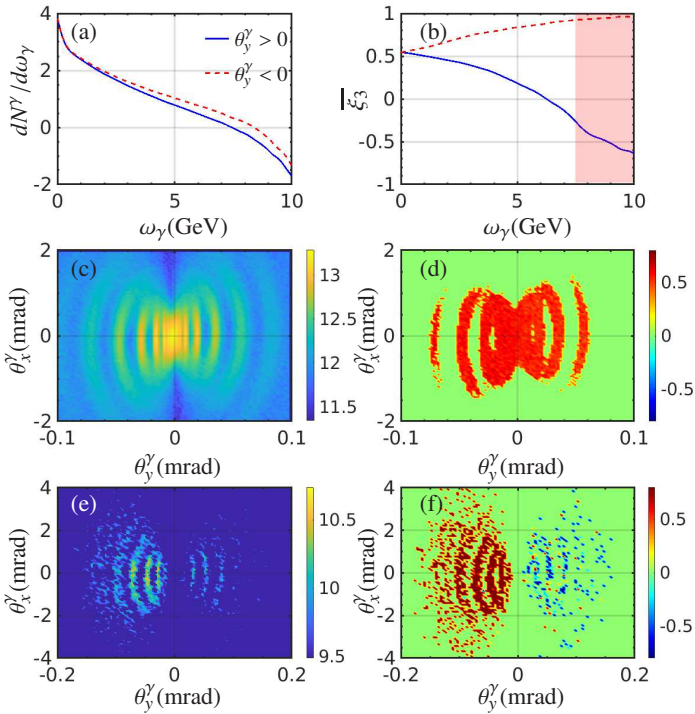}
	\begin{picture}(300,20)
	\put(18,267){(a)}
	\put(150,267){(b)}
	\put(18,177){(c)}
	\put(150,177){(d)}
	\put(18,88){{\color{white}(e)}}
	\put(150,88){(f)}
	\put(42,15){$\theta^\gamma_y$(mrad)}
	\put(172,15){$\theta^\gamma_y$(mrad)}
	\put(42,102){$\theta^\gamma_y$(mrad)}
	\put(172,102){$\theta^\gamma_y$(mrad)}
    \put(44,193){$\omega_\gamma$(GeV)}
	\put(174,193){$\omega_\gamma$(GeV)}
	\put(-7,140){\rotatebox{90}{$\theta^\gamma_x$(mrad)}}
    \put(-5,230){\rotatebox{90}{$dN^\gamma/d\omega_\gamma$}}
    \put(125,140){\rotatebox{90}{$\theta^\gamma_x$(mrad)}}
    \put(120,240){\rotatebox{90}{$\overline{\xi_3}$}}
    \put(-7,49){\rotatebox{90}{$\theta^\gamma_x$(mrad)}}
    \put(125,49){\rotatebox{90}{$\theta^\gamma_x$(mrad)}}
    \put(80,264){\footnotesize $\theta^\gamma_y>0$}
    \put(80,251){\footnotesize $\theta^\gamma_y<0$}
	\end{picture}
	\caption{ (top row): (a) log$_{10}dN^\gamma/d\omega_\gamma$, and (b) averaged Stokes parameter $\overline{\xi_3}$ of gamma-photons emitted in $\theta^\gamma_y>0$ (solid line) and $\theta^\gamma_y<0$ (dashed line) vs photon energy $\omega_\gamma$ in $|\theta_x|,|\theta_y|<10$ mrad. (middle row): (c) Angular distribution of $d^2N^\gamma/d\theta^\gamma_xd\theta^\gamma_y$ (rad$^{-2}$), and (d) $\overline{\xi_3}$ vs $\theta^\gamma_x=k_x/k_z$ and $\theta^\gamma_y=k_y/k_z$. (bottom row): (e) Angular distribution of $d^2N^\gamma/d\theta^\gamma_xd\theta^\gamma_y$ (rad$^{-2}$), and (f) $\overline{\xi_3}$ for photons with energy $\varepsilon>7.5$ GeV [shaded red  in (b)]).    }
	\label{Fig. 4}
\end{figure}

To understand the effects of photon polarization on pair production, we investigate the polarization of emitted photons in $\theta^\gamma_y>0$ and $\theta^\gamma_y<0$ separately, as shown in Fig. \ref{Fig. 4}. The photon density emitted within the angular region $\theta^\gamma_y<0$ is larger than at $\theta^\gamma_y>0$, especially in high energy region, as shown in Fig. \ref{Fig. 4}(a). When an electron with $\vec{\zeta}_i=\vec{b}$ counterpropagate to the laser field, the direction of the instantaneous quantization axis is $\vec{n}=\vec{\zeta}^{f}/|\vec{\zeta}^{f}|$\cite{li2020production} with
\begin{equation}
\vec{\zeta}^{f}=\frac{\left(2f_{2}-f_{1}\right)\vec{\zeta}_i-\frac{\omega}{\varepsilon'}\vec{b}f_{3}+\frac{\omega}{\varepsilon'\varepsilon}\left[f_{2}-f_{1}\right]\left(\vec{\zeta}_i\vec{\hat{v}}\right)\vec{\hat{v}}}{\frac{\varepsilon^{2}+\varepsilon'^{2}}{\varepsilon'\varepsilon}f_{2}-f_{1}-\frac{\omega}{\varepsilon}\vec{\zeta_i}\vec{b}f_{3}},
\end{equation}
which is mostly along $B_y$, and changes sign every half-cycle.
For the electron with initial spin $\zeta_y=1$, its spin is parallel (spin up) and anti-parallel (spin down) to the quantization axis in the half-cycles with $B_y>0$ and $B_y<0$, respectively. The emission probability is larger when electron is spin down before the emission, as shown in Fig.~\ref{Fig. 4}(a), which is in accordance with the analysis of the spin resolved probability given in Fig.~\ref{Fig. 1}(a). Further, the photon emission direction is parallel to the momentum of the emitting particle, and the photons emitted in $B_y>0$ ($B_y<0$) propagate with $p_y<0$ ($p_y>0$), respectively, because the oscillation phase of $p_y$ has a $\pi$-delay with respect to $B_y$.
%Meanwhile, since and the emission direction is parallel with the momentum of the emitting particle, the photons emitted in $B_y>0$ and $B_y<0$ propagate with $p_y<0$ and $p_y>0$, respectively.
Therefore, the electrons emit photons with $\theta^\gamma_y>0$ at $B_y>0$ and emit photons with $\theta^\gamma_y<0$ at $B_y<0$. The latter has higher emitted photon number than former due to the larger emission probability $W_{r\downarrow}>W_{r\uparrow}$, as explained above.
More importantly, the electrons with spin-up (spin-down) have a higher probability to emit photons with $-1<\xi_3<0.5$ ($0.5<\xi_3<1$). Therefore, the radiation with $\theta_y>0$ mainly comes from photon emission at $B_y>0$, and has polarization $-1<\xi_3<0.5$. While the radiation at $\theta_y<0$ comes from photon emission at $B_y<0$, which has polarization $0.5<\xi_3<1$, as shown in Fig.~\ref{Fig. 4}(b).

The fringes in the angular distribution, seen in Figs.~~\ref{Fig. 4}(c)-(e), are due to the radiation in different laser cycles.
%The polarization of photons emit in $\theta_y^\gamma<0$ is slightly larger than that in $\theta_y^\gamma>0$.
As shown in Fig.~\ref{Fig. 4}(b), $\overline{\xi}_3$ in high energy region is positive for spin down electrons while negative for spin up electrons.  However, since the radiation is dominated by low energy photons with $\xi_3\sim 0.55$, the angular distribution of the photon polarization in Fig.~\ref{Fig. 4}(d) is also dominated by $\xi_3\sim 0.55$.
%the polarization of positrons $\overline{\xi_3}\sim 0.55$ for both  $\theta^\gamma_y>0$  and  $\theta^\gamma_y<0$ , as shown in Fig.~\ref{Fig. 4}(d).  
Nevertheless, after filtering the low energy emissions, the correlation of photon polarization and emission angle can be seen in Figs. \ref{Fig. 4}(e)  and (f). As expected, the photons distributed in $\theta^\gamma_y<0$ have $\overline{\xi_3}<0$, while in the region $\theta^\gamma_y<0$ the polarization is $\overline{\xi_3}>0$. Since the pairs are produced mostly by energetic photons, 
%are more probably to produce pairs, 
the distinct polarization properties of high energy photons in $\theta^\gamma_y<0$ and $\theta^\gamma_y>0$ break the symmetric angular distribution of polarization.

The separation of  positron polarization along propagation direction can be explained\cite{li2019ultrarelativistic}, taking into account that the final momentum of the created electron (positron) is determined by the laser vector-potential $A_y(t_p)$ at the creation moment $t_p$: $p_f=p_i+eA_y(t_p)$, where 
%$t_p$ is the creation time,  $A_y(t_p)$ the vector potential at  $t_p$, $p_f$ the final momentum of positron, and  
$p_i$ is the momentum inherited from parent photon, with the vanishing average value $\overline{p}_i=0$.
%which is an arbitrary value with $\overline{p}_i=0$.
On the other hand, when the photon 
%polarization of the emitted photon 
is linearly polarized with (0,0,$\xi_3$), the SQA for pair production of such a photon is along the laser magnetic field direction, $\vec{n}=\vec{\zeta}_+^{f,\xi}/|\vec{\zeta}_+^{f,\xi}|=\vec{b}$. Therefore,
the positrons produced at  $A_y(t_p)<0$ obtain a final momentum  $p_f\approx eA_y(t_p)>0$ and is polarized along  $\zeta_y>0$, as the instantaneous SQA is along $y>0$ at $t_p$. Similarly, one has $\zeta_y<0$ when  $p_f<0$.

\begin{figure}
	\setlength{\abovecaptionskip}{-0.2cm}
	\includegraphics[width=0.48\textwidth]{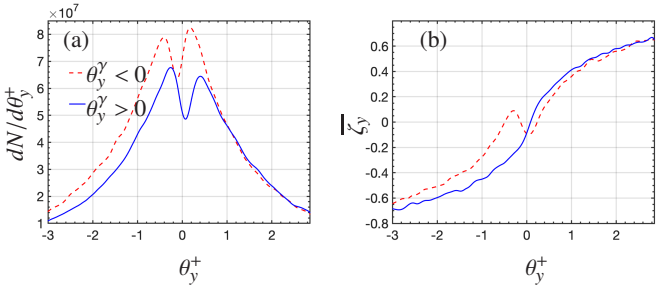}
	\begin{picture}(300,20)
	\put(15,98){(a)}
	\put(150,98){(b)}
	\put(24,85){$\theta^\gamma_y<0$}
	\put(24,72){$\theta^\gamma_y>0$}
	\put(59,12){$\theta^+_y$}
	\put(190,12){$\theta^+_y$}
    \put(-7,55){\rotatebox{90}{$dN/d\theta^+_y$}}
    \put(120,65){\rotatebox{90}{$\overline{\zeta_y}$}}
  %  \put(186,67){{\color{red}$\bullet$}}
  %  \put(190,67){{\color{blue}$\bullet$}}
   %  \put(56,85){{\color{red}$\bullet$}}
   % \put(59.5,69.5){{\color{blue}$\bullet$}}
	\end{picture}
	\caption{(a) Polarized positron density distribution d$N/$d$\theta_y^+$ (rad$^{-1}$) vs positron polar angle $\theta_y^+$ (rad) for positrons produced by photon with $\theta^\gamma_y>0$ (solid line) and $\theta^\gamma_y<0$ (dotted line); (b) Averaged polarization degree $\overline{\zeta_y}$ versus $\theta_y$ (rad) for positrons produced by photon with $\theta^\gamma_y>0$ (solid line) and $\theta^\gamma_y<0$ (dotted line). }
	\label{Fig. 5}
\end{figure}

To reveal the origin of the abnormal polarization features in a small angle region, we artificially turn on pair production of  photons with $\theta_y^\gamma>0$ and $\theta_y^\gamma<0$ separately.
As shown in Fig.~\ref{Fig. 4}(b), the photons with $\theta_y^\gamma<0$ have $\xi_3\sim1$  at high energy end of spectrum. Those hard photons have a higher probability to produce energetic positrons anti-parallel with magnetic field, as discussed in  Sec.~I.B, resulting in a reverse of polarization direction in small angle region and a overall decrease of averaged polarization of positrons, as shown in Fig.~\ref{Fig. 5}(b). Moreover, since  the parent photon with  $\theta_y^\gamma<0$ have $\overline{p}_i<0$, the positrons distribution is slightly shifted towards  $\theta_y^+<0$.
Meanwhile, the polarization of hard photons at $\theta_y^\gamma>0$ side  is reaching to $\xi_3\sim-1$ [Fig.~\ref{Fig. 4}(b)]. If these photons are collected to produce pairs, highly polarized positrons could be achieved due to the highly domination of the spin-up positrons, $dW_{\downarrow\uparrow}(\xi_3=-1)\gg dW_{\downarrow\uparrow}(\xi_3=0)$. In the present case, the $\overline{\zeta}_y$ of positrons produced by photons with  $\theta_y^\gamma>0$ increases monotonously with $\theta_y^+$, and the angular distribution shift slightly towards $\theta_y^+>0$.
Since the pair production probability  is inversely proportional to $\xi_3$, and the photons with $\theta_y^\gamma<0$ have larger $\overline{\xi_3}$ than those with  $\theta_y^\gamma>0$, therefore, more positrons are produced at $\theta_y^\gamma>0$, as shown in Fig.~\ref{Fig. 5}(a). Thus, the domination of pair production of photons in $\theta_y^\gamma<0$ and $\overline{p}_i<0$ result in a decrease of the polarization of the positron beam and an asymmetric angular distribution of polarization.

%Moreover, for positrons produced by photons with $\theta_y^\gamma<0$  ($\theta_y^\gamma<0$), the symmetric centres of angular distribution of positron density and polarization shift slightly towards $\theta_y<0$  ($\theta_y>0$), due to the contribution of $p_i$ inherited from parent photons, as shown in Fig. \ref{Fig. 5}. The contributions from both sides result in a shift of polarization zero in Fig. \ref{Fig. 3} (f).

\section{Conclusion}

We have investigated the photon polarization  effects on pair production in nonlinear Breit-Wheeler process via a newly developed Monte Carlo method employing the fully polarization resolved quantum probabilities. We show that the longitudinal polarization of produced positrons is solely induced by the photon polarization.
%The exclusion of photon polarization gets rid of the longitudinal polarization of produced positrons.
While the transverse polarization of positrons could either increase, decrease or even be unchanged determined by the polarization of intermediate gamma-photons. For the interaction of initially transversely polarized electrons and an elliptically polarized laser, both the polarization degree and density of positrons are reduced when the polarization of the intermediate photons are accounted for. This is because the photons emitted during nonlinear Compton process are partially   polarized along electric field direction with $\overline{\xi_3}\approx 0.55$. The hard photons in $\theta_y^\gamma<0$ have even higher polarization $\overline{\xi_3}\sim 1$, causing the energetic positrons produced in small angle region reverse the polarization direction. If one separates the intermediate hard gamma-photons within $\theta_y^\gamma>0$, the polarization of positrons will be highly enhanced due to the domination of $dW_{\downarrow\uparrow}$ probabilities through the spectrum. Our results confirm that the important role of the intermediate photon polarization
%modified the numerical method of describing polarization 
during strong-field QED process and should be accounted for in designing and optimizing a realistic laser driven polarized positron source. Moreover, the measurement of the correlated electron and positron polarization in pair production process can shed light on the intermediate interaction dynamics, in particular, on the polarization properties of intermediate photons. 
%The investigations of  the  photon polarization effects on positrons provide new ways towards improving positrons polarization and  detecting gamma-photon polarization, which is of great interest for high energy physics and astrophysics.

\bibliography{prp}

\end{document}